\documentclass[%
 reprint,
 amsmath,amssymb,
 aps,
]{revtex4-1}

\usepackage{graphicx}
\usepackage{dcolumn}
\usepackage{bm}
\usepackage{subfigure}
\usepackage{booktabs}

\begin{document}


\title{The determination of $\alpha$-spectroscopic factors and ANC of $^{16}$O states using $^{12}$C($^{20}$Ne,$^{16}$O)$^{16}$O reaction at $E_{lab}$=150 MeV incident energy}

\author{Ashok Kumar Mondal$^1$}
\author{C. Basu$^{1}$}
\email{drchinmaybasu@gmail.com}
\author{S. Adhikari$^2$}
\author{C. Bhattacharya$^3$}
\author{T. K. Rana$^3$}
\author{S. Kundu$^3$}
\author{S. Manna$^3$}
\author{R. Pandey$^3$}
\author{P. Roy$^3$}
\author{ A. Sen$^3$}
\author{J. K. Meena$^3$}
\author{A. K. Saha$^3$}
\author{J. K. Sahoo$^3$}
\author{Dipali Basak$^1$}
\author{Tanmay Bar$^1$}
\author{Haridas Pai$^1$}
\author{A. Bisoi$^4$}
\author{A. K. Mitra$^1$}
\author{Piyasi Biswas$^1$}
 \affiliation{$^1$Nuclear Physics Division, Saha Institute Of Nuclear Physics, HBNI, 1/AF Bidhan Nagar, Kolkata- 700064, INDIA.}
 \affiliation{$^2$Techno India University, EM-4/1, Sector-V, Salt Lake, Kolkata-700091, INDIA.}
 \affiliation{$^3$Variable Energy Cyclotron Centre, Kolkata - 700064, INDIA.}
 \affiliation{$^4$Indian Institute of Engineering Science and Technology, Shibpur, Howrah -711103, INDIA.}

\date{\today}

\begin{abstract}
The $^{12}$C($^{20}$Ne,$^{16}$O)$^{16}$O $\alpha$-transfer reaction at $E_{lab}$=150 MeV is first time used to determine the ANC of the 6.92 MeV and 7.12 MeV states of $^{16}$O. The $^{20}$Ne+$^{12}$C potential parameters are also obtained from elastic scattering. The direct reaction code FRESCO is used to determine the $\alpha$-spectroscopy factor ($S_{\alpha}$) of the three states of $^{16}$O (6.92 MeV, 7.12 MeV and 11.52 MeV) and ANC of the two states (6.92 MeV and 7.12 MeV) of $^{16}$O. The extracted ANC and $S_{\alpha}$ are compareable to previous measurements.
\end{abstract}

\pacs{Valid PACS appear here}
\maketitle


\section{Introduction}

The Asymptotic Normalization Co-efficient (ANC) method [1,2] is an indirect method to study astrophysical reactions at low energies. Such reactions take place in stars at relative energies much below the Coulomb barrier of the two interacting nuclei (except in neutron induced reactions). As a result, the cross-sections are very small and their direct measurement with reasonable accuracy is very difficult or almost impossible with presently available techniques. R-matrix extrapolation [3] of the cross-section measured at higher energy to the Gamow energy is the solution of this problem. Additionally, if the main component of the capture cross-section is external capture then the determination of the ANC can evalute the S-factor at zero energy.
The $^{12}$C($\alpha$,$\gamma$) reaction is such a reaction for which the ANC method is most suitable [4].  The rate of this reaction greatly affects the resulting ratio of $^{12}$C to $^{16}$O and also the final fate of the star (i.e., black hole or neutron star). At 300 keV (Gamow energy) the cross-section of this reaction is of the order of $10^{-17}$ barn and its direct measurement is almost impossible. The only method to determine the $^{12}$C($\alpha$,$\gamma$)$^{16}$O reaction cross-section at 300 keV is by R-matrix extrapolation . The extrapolation relies on the $\alpha$-spectroscopic properties of $^{16}$O states [5,6,7,8,9,10]. One can allow these parameters to vary freely in the extrapolation process. But a more physical away is to determine these parameters from some indirect method such as the ANC technique. The alpha capture in the $^{12}$C($\alpha$,$\gamma$)$^{16}$O reaction is an external capture process and proceeds through mainly two sub-threshold states (6.92 MeV (2$^{+}$) and  7.12 MeV (1$^{-}$)) of $^{16}$O. The ANC of these two sub-threshold states play a crucial role in determination of the capture cross-section. The $^{12}$C($^{6}$Li,d) and $^{12}$C($^{7}$Li,t) $\alpha$-transfer reactions [2, 5, 7, 11, 12] have been mainly used to determine the ANC at both below and above Coulomb barrier energies. Besides, $^{12}$C($^{11}$B,$^{7}$Li)$^{16}$O reaction has been also used for the same purpose [13]. Though, at sub-Coulomb energeis the determination of ANC is more model independent, the above barrier measurements are more convenient due to larger cross-sections. 

In this work, a new reaction $^{12}$C($^{20}$Ne,$^{16}$O)$^{16}$O is used for the first time to determine the $S_{\alpha}$ and ANC of $^{16}$O states and using this ANC of $^{16}$O states we have extracted $S_{E2}$ value at 300 keV by R-matrix extrapolation. The measurements are carried out at an incident energy of 150 MeV.

\section{Experimental Details}
The experiment was carried out at the K130 room temperature Cyclotron centre at VECC, Kolkata. The 1 m  scattering chamber at channel 2 was used for the measurements. Three \textit{$\Delta$E-E} telescopes were setup inside the chamber. One telescope consisted of \textit{$\Delta$E-E} setup with (50 mm x 50 mm) silicon strip detectors. The \textit{$\Delta$E} strip detector was 50 $\mu$m thick and the \textit{E} strip ( double sided) 500 $\mu$m. The other telescopes (\textit{T1} and \textit{T2}) were  15 $\mu$m- 1000 $\mu$m surface barrier detectors. The target was self supporting 660 $\mu$gm/$cm^2$ $^{12}$C foils. A gold target was mounted in the ladder for energy calibration purpose. The beam current was stable at around 10-20 pnA.There were 16 strips in the \textit{$\Delta$E} detector with 100 $\mu$m dead layer between adjacent strips. The centre to centre distance between adjacent strip was 3 mm. The strip telescope covered angles from 14.82$^{o}$ to 34.18$^{o}$ whereas the low threshold surface barrier telescopes were used to measure the cross-sections at backward angles from 30$^{o}$ to 52$^{o}$. Typical energy resolutions of the \textit{$\Delta$E} and \textit{E} strip detectors were 40 keV and 60 keV respectively. The solid angles subtended by the telescopes at the target centre were 0.796 msr(strip detector),0.128 msr(\textit{T1} telescope) and 0.2716 msr (\textit{T2} telescope) respectively. The $^{20}$Ne + $^{12}$C reaction have been measured in the angular range ${15}^o$ to ${52}^o$ at $E_{lab}$=150 MeV energy. The distance between target and strip detector \textit{$\Delta$E-E} telescopes was 184.15 mm and with surface barrier telescopes were 235 mm. The dimension of the collimator in the strip detector was 50 mm x 9 mm (rectangular) and for \textit{T1} and \textit{T2}, 3 mm diameter circular and 3 mm x 5 mm rectangular respectively. VME data aquisition system was used in the experiment.

\section{Results and Discussion}

In this experiment $^{20}{Ne}^{10^{+}}$ beam of energy $E_{lab}$=150 MeV was incident on a $^{12}$C target and the 2-D spectrum for a representative case of 19.82$^{o}$ for the strip telescope is shown in fig.1 and that for the telescope \textit{T1} at 36$^{o}$ is shown in fig.2.
\begin{figure}[!htbp]
\centering
\includegraphics[scale=0.26]{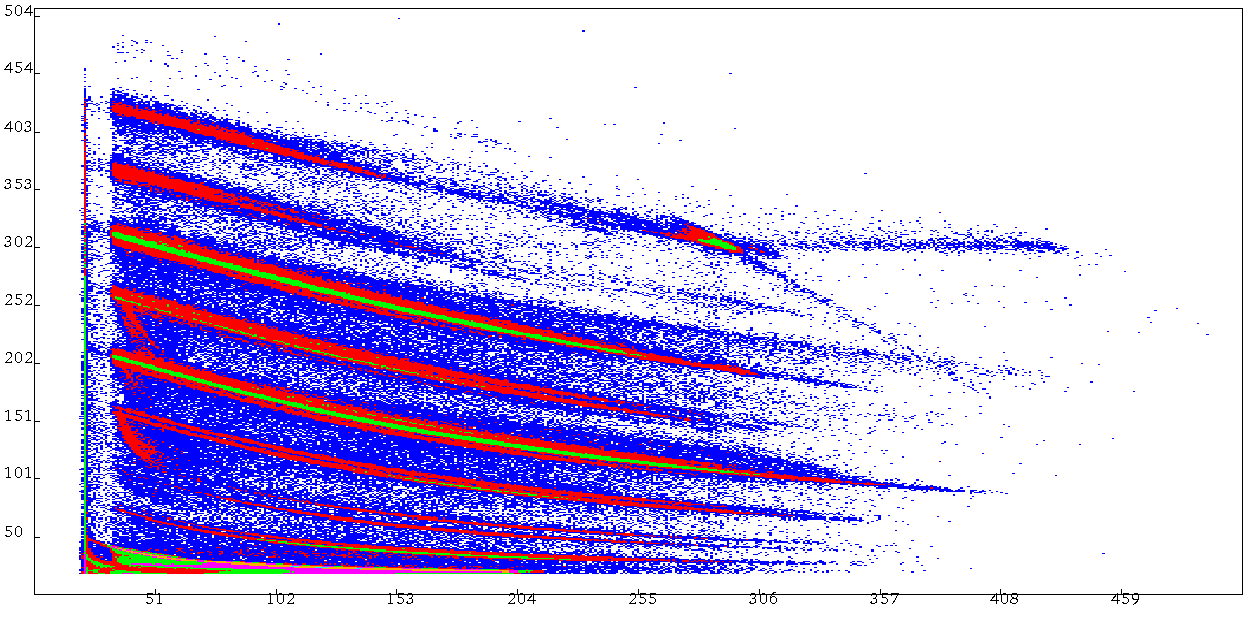}
\caption{\label{fig.1}{\scriptsize 2-D energy spectrum of the strip detector \textit{$\Delta$E-E} telescope for the $^{20}$Ne+$^{12}$C system at $\Theta_{lab}$=19.82$^o$ for E$^{lab}_{^{20}Ne}$=150 MeV.}}
\end{figure}

\begin{figure}[!htbp]
\centering
\includegraphics[scale=0.26]{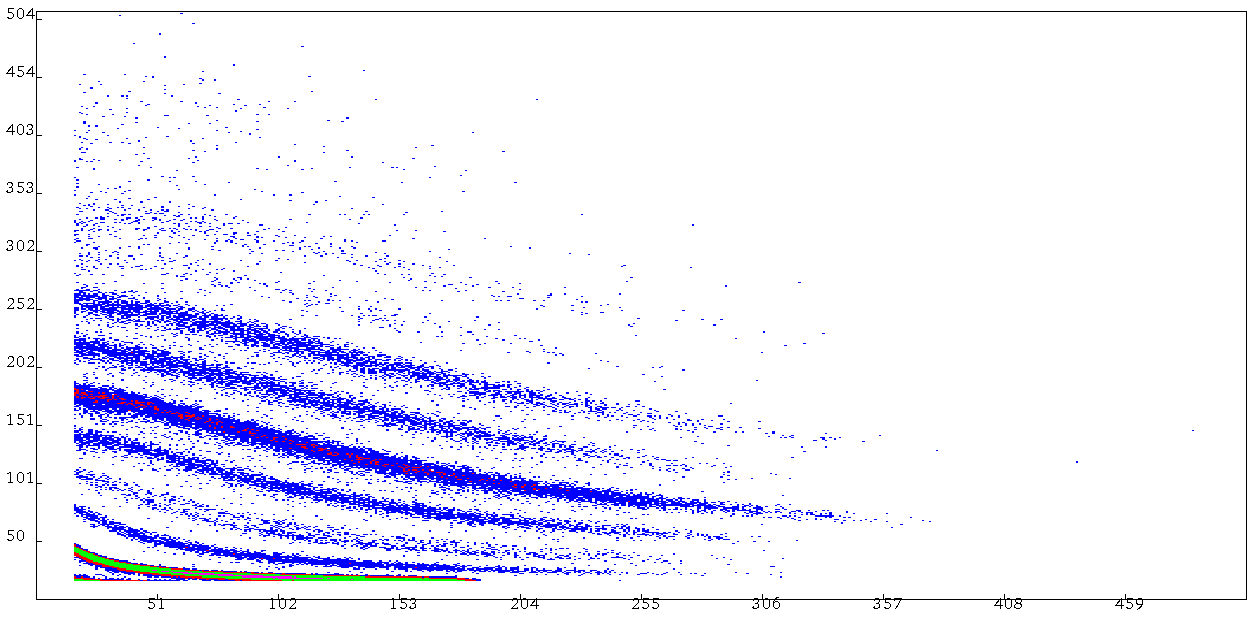}
\caption{\label{fig.2}{\scriptsize  2-D energy spectrum of the SB detector \textit{$\Delta$E-E} telescope T1 for the $^{20}$Ne+$^{12}$C system at $\Theta_{lab}$=36$^o$ for E$^{lab}_{^{20}Ne}$=150 MeV.}}
\end{figure}

\begin{figure}[!htbp]
\centering
\includegraphics[scale=0.55]{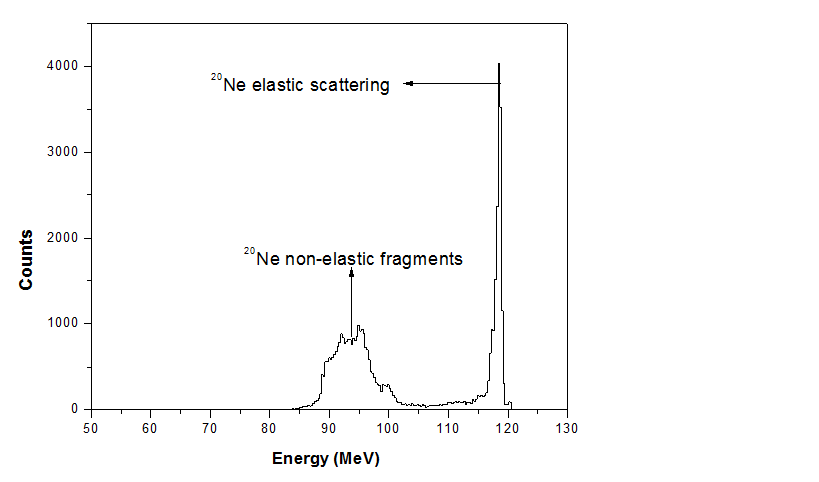}
\caption{\label{fig.3} {\scriptsize Energy spectrum of elastic scattering of $^{20}$Ne from $^{12}$C at $\Theta_{lab}$=19.82$^o$ for E$^{lab}_{^{20}Ne}$=150 MeV using strip detector \textit{$\Delta$E-E} telescope. Here at higher energies there is the strong elastic peak of $^{20}$Ne and at lower energies there are $^{20}$Ne nonelastic fragments. }}
\end{figure}

\begin{figure*}[!htbp]
\centering
  \mbox{
    \subfigure[$\Theta_{lab}$=19.82$^o$]{
      \includegraphics[width=0.50\textwidth]{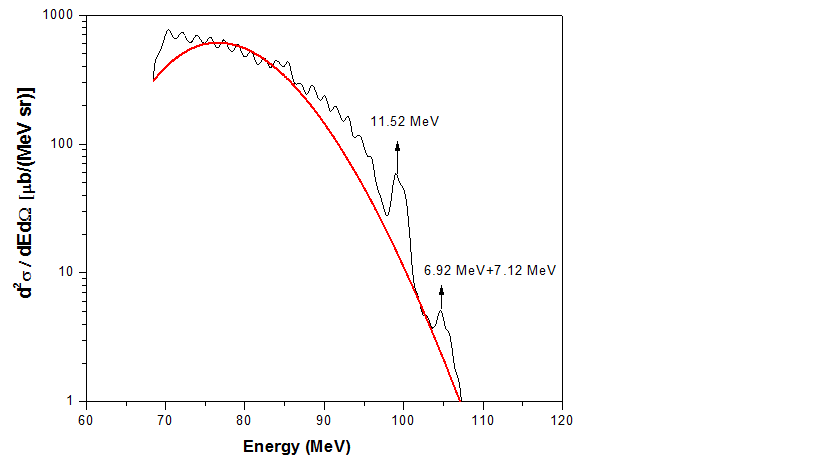}
}
\subfigure[$\Theta_{lab}$=20.77$^o$]{
  \includegraphics[width=0.50\textwidth]{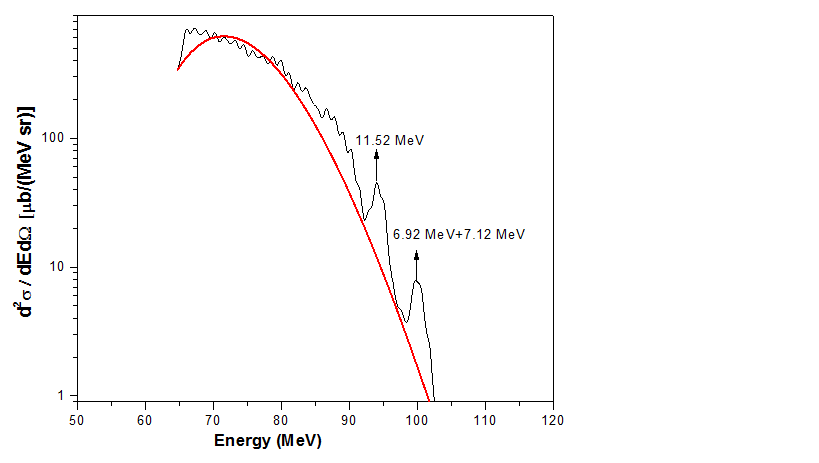}
}} \\ 
\mbox{
\subfigure[$\Theta_{lab}$=21.73$^o$]{
  \includegraphics[width=0.50\textwidth]{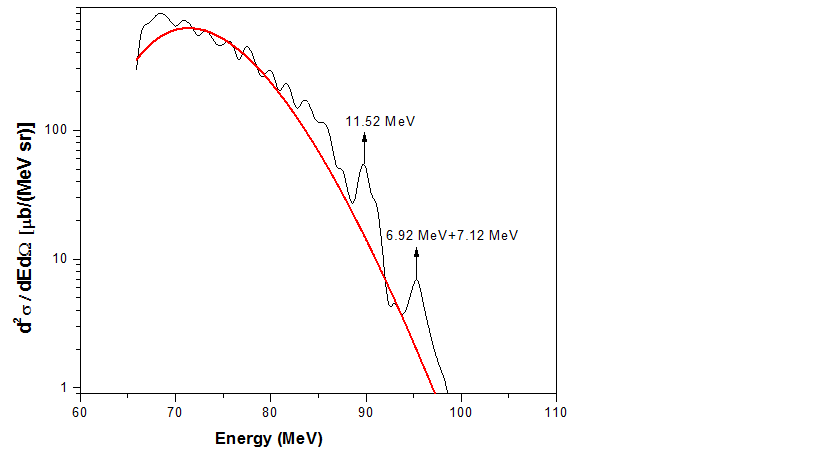}
}%
\subfigure[$\Theta_{lab}$=22.68$^o$]{
  \includegraphics[width=0.50\textwidth]{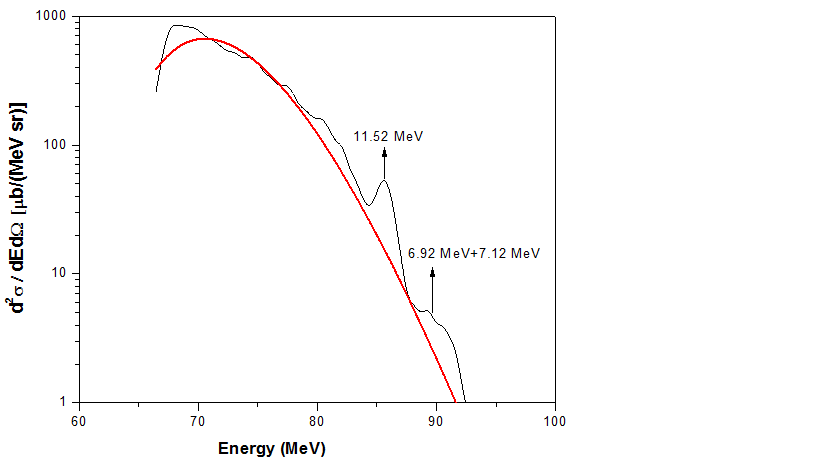}
}} \\ 
\mbox{
\subfigure[$\Theta_{lab}$=23.64$^o$]{
   \includegraphics[width=0.50\textwidth]{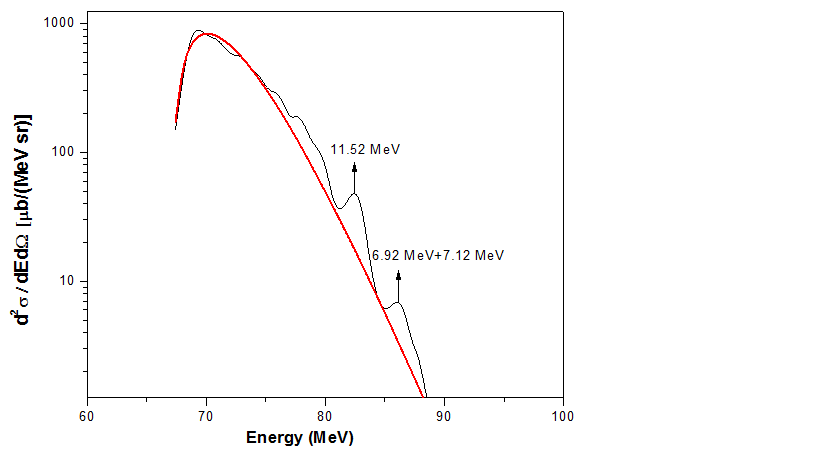}
}}
\caption{ {\scriptsize Energy spectrum of the $^{16}$O in strip detector \textit{$\Delta$E-E} telescope for the $^{12}$C($^{20}$Ne,$^{16}$O)$^{16}$O reaction at different angles for E$^{lab}_{^{20}Ne}$=150 MeV which are fitted by moving source model [14]. The red line fit the low energy continuous background using moving source model.}}
\label{subfig}
\end{figure*}

Emitted fragments (2$\geq$Z$\leq$10) are clearly seperated as seen in these plots. Fig.3 shows the projected elastic and nonelastic fragments at the angle $\Theta_{lab}$=19.82$^o$. Gating on the Z=8 curve from the 2-D spectrum in fig.1 and projecting on the energy axis,  we obtain the energy spectrum of the emitted $^{16}$O fragments. These spectra at various angles are shown in fig.4. The spectrum has a broad continuous low energy section and some discrete peaks at higher energy. These discrete peaks can results from transfer reactions or compound emissions populating discrete low lying states of $^{16}$O. The three discrete peaks were identified at 6.92 MeV, 7.12 MeV and 11.52 MeV. In order to obtain the cross-sections of these three states, the cross-section of the continuous part has to be estimated as they contribute to the background of the discrete peaks.

The shape of the continuous part suggest a statistical process. In order to calculate the cross-section for emission of such heavy fragments (Z=8) from statistical process, we used the moving source model [14]. The parameters in this model are $N_{0}$, \textit{Z$E_{C}$}, $E_{1}$, and \textit{T}, Where $N_{0}$ is an overall normalization constant, $E_{C}$ is the kinetic energy gained by the light particle of charge \textit{Z} due to the Coulomb repulsion from the target, $E_{1}$ is the kinetic energy of a particle at rest in the moving frame and T is the source temperature. The value of these parameters required to calculate the $^{16}$O spectrum at various angles are given in the Table I. The cross-sections of the discrete peaks are estimated by subtracting the background calculated from moving source model and are plotted in fig.7.

\begin{figure*}[!htbp]
\centering
  \mbox{
    \subfigure[$\Theta_{lab}$=19.82$^o$]{
      \includegraphics[width=0.50\textwidth]{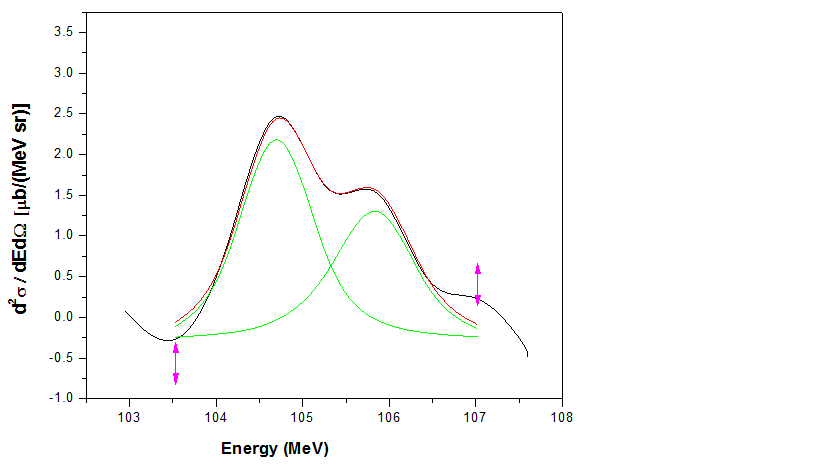}
}
\subfigure[$\Theta_{lab}$=20.77$^o$]{
  \includegraphics[width=0.50\textwidth]{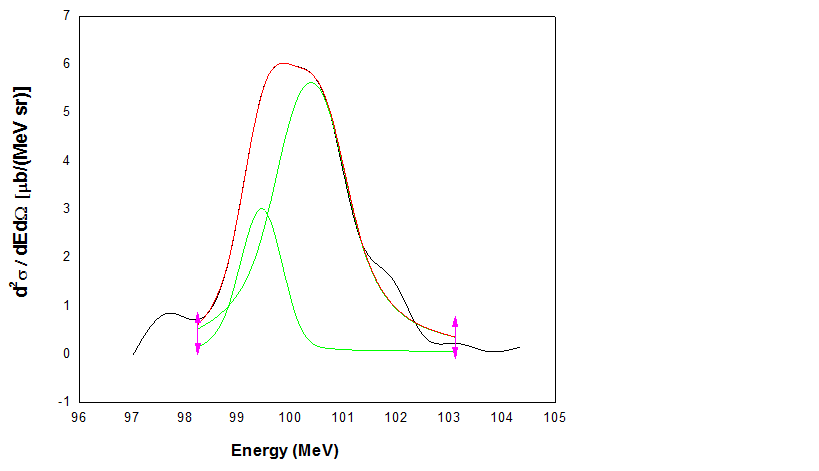}
}} \\ 
\mbox{
\subfigure[$\Theta_{lab}$=21.73$^o$]{
  \includegraphics[width=0.50\textwidth]{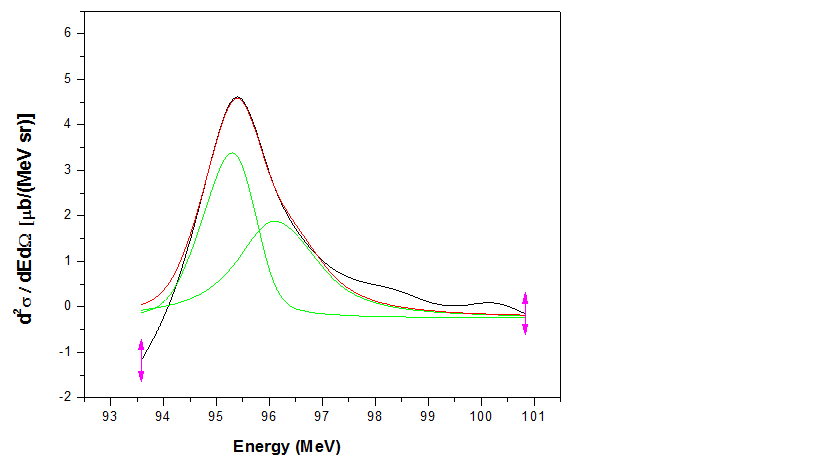}
}%
\subfigure[$\Theta_{lab}$=22.68$^o$]{
  \includegraphics[width=0.50\textwidth]{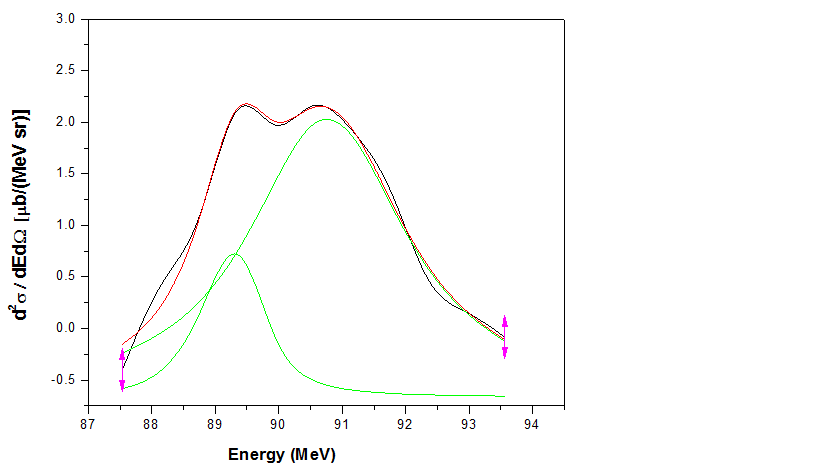}
}} \\ 
\mbox{
\subfigure[$\Theta_{lab}$=23.64$^o$]{
   \includegraphics[width=0.50\textwidth]{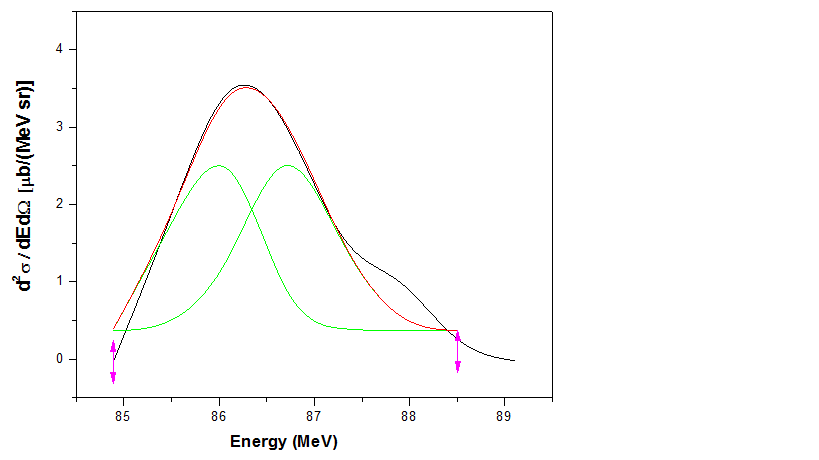}
}}
\caption{ {\scriptsize Fitting of the 6.92 MeV and 7.12 MeV peaks of the $^{16}$O after substracting the  low energy continuous background by moving source model [14] at different angles.}}
\label{subfig}
\end{figure*}

\begin{table*}[!htbp]
\caption{The parameters in moving source model at various angles are given below.} 
\centering 
\small
\begin{tabular}{c  c  c  c  c} 
\hline\hline  \\
Lab angle  & $N_{0}$   & \textit{Z}$E_{C}$ (MeV)   & $E_{1}$ (MeV)   & \textit{T} (MeV) \\ [2ex]  \\
\hline  \\
19.82 & 239786 $\pm$ 86754 & 36.15 $\pm$ 0.173  & 45 & 0.677 $\pm$ 0.0301  \\ [2ex]  
20.77 & 4581 $\pm$ 833 & 46.19 $\pm$ 0.166  & 28 & 0.91 $\pm$ 0.0423  \\ [2ex]
21.73 & 3322 $\pm$ 54 & 53.25 $\pm$ 0.126  & 20 & 0.9 $\pm$ 0.035  \\ [2ex]
22.68 & 2025 $\pm$ 30 & 58.67 $\pm$ 0.0912  & 13 & 0.85 $\pm$ 0.032  \\ [2ex]
23.64 & 696 $\pm$ 9 & 67.19 $\pm$ 0.0.036  & 1.89 $\pm$ 0.56 & 1.45 $\pm$ 0.052  \\ [2ex]
\hline 
\end{tabular}
\label{table:nonlin} 
\end{table*}

Indirect method uses transfer reactions to determine the  ANC of the bound states of the nuclei of interest. In this work, the transfer reaction ( $^{12}$C($^{20}$Ne,$^{16}$O)$^{16}$O) is chosen to produce  $^{16}$O through $\alpha$-transfer, the ANC of whose states have to be determined. The measured cross-sections are then compared to a single particle transfer reaction model such as the DWBA (Distorted Wave Born Approximation)[1] theory. The $\alpha$-spectroscopic factor ($S_{\alpha}$) of a nuclear state can be extracted from $\alpha$-transfer reaction by a normalization of the experimental data with the theoretical cross-section. Thus,
 \begin{equation}
 ( \frac{d\sigma}{d\Omega} )_{Expt} = S_1 S_{\alpha} ( \frac{d\sigma}{d\Omega} )_{Theo}
 \end{equation} 
 
where $S_1$ is the spectroscopic factor for the $\alpha$ + $^{16}$O configuration of the $^{20}$Ne ground state and $S_{\alpha}$ is the $\alpha$ + $^{12}$C spectroscopic factor for a state of $^{16}$O. The square of the ANC (C$^2$) of a particular state is related to the alpha spectroscopic factor ($S_{\alpha}$) via the single particle ANC b$^{2}$ as 
\begin{equation}
C^2 = S_{\alpha} b^2
\end{equation}
The single particle ANC $\textit{b}$ is the normalization of the bound state wave function of $^{16}$O at large radii with respect to the Whittaker function and is calculated from a suitable binding potential. \\

In fig.7, the measured $^{16}$O angular distribution (symbol) is shown for the 6.92 MeV, 7.12 MeV and 11.52 MeV states.
 The calculated cross-sections (solid lines) are obtained by using the code FRESCO [15] (version fres2.9) in the framework of the DWBA theory. The required exit channel ($^{16}$O + $^{16}$O) potential and the core-core $^{16}$O + $^{12}$C potential are obtained respectively from [6] and [16].  The entrance channel potential are extracted from elastic scattering data measured in the present work. The $\alpha$ + $^{16}$O and $\alpha$ + $^{12}$C binding real potentials have a Gaussian form factor. The depth of the potential is adjusted to fit the respective seperation energies of $\alpha$ in $^{20}$Ne and $^{16}$O. 
 The optical model analysis for the elastic scattering data is performed with a Woods-Saxon (WS) potential i.e  
\[U(r)=\frac{-V_o}{1+exp(\frac{r-R_o}{a_o})}+\frac{-iW}{1+exp(\frac{r-R_w}{a_w})}\]
where $V_o$, $R_o$ and $a_o$ are real potential parameters and W, $R_w$ and $a_w$ are imaginary potential parameters (Where $R_c$= 1.2$A_T^{1/3}$).\\
Elastic scattering of $^{20}$Ne on $^{12}$C at $E_{lab}$=150 MeV from  $40^o$ to $65^o$ in the C.M system is used to extract the potential parameters for the $^{20}$Ne+$^{12}$C system. The open black square in fig.6 are the experimental elastic scattering data measured in the present work and dotted red line is the SFRESCO fit [15]. The extracted potential parameters are shown in Table.2. This parameters are used in $S_{\alpha}$ and the calculation of $^{12}$C($^{20}$Ne,$^{16}$O)$^{16}$O transfer cross-section for the ANC  of $^{16}$O states.

\begin{figure}[!htbp]
\centering
\includegraphics[scale=0.5]{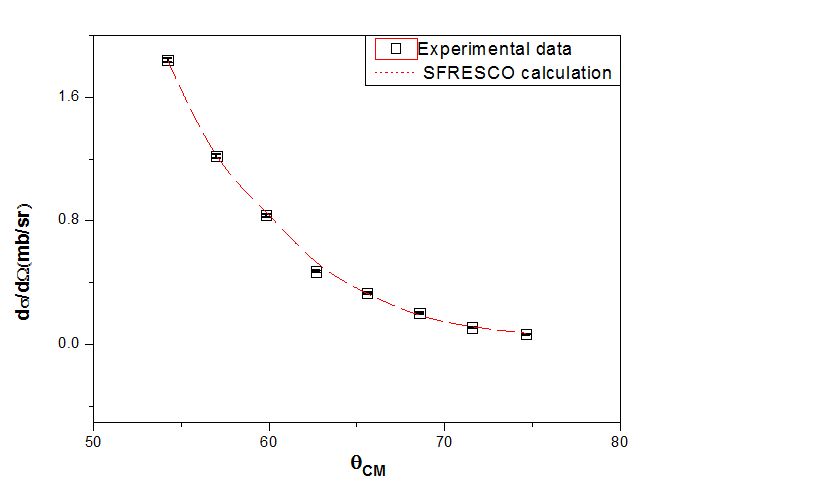}
\caption{\label{fig.4}{\scriptsize Angular distribution of $^{12}$C($^{20}$Ne,$^{20}$Ne)$^{12}$C elastic scattering cross-section at $E_lab$=150 MeV from  $40^o$ to $65^o$ in C.M system. Here open black square  are experimental elastic scattering cross-section data with error and dotted red line is the SFRESCO calculation.}}
\end{figure}

\begin{figure*}[!htbp]
\centering
\mbox{
\subfigure[ 6.92 MeV state of $^{16}$O ]{
 \includegraphics[scale=0.45]{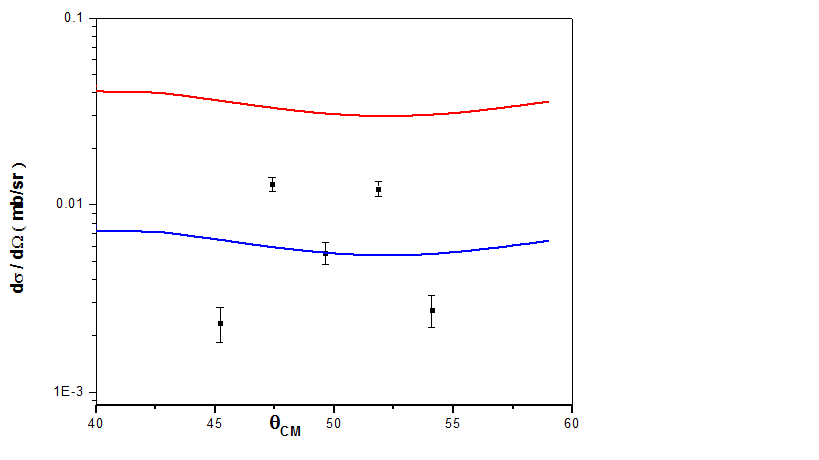}
 }
\subfigure[ 7.12 MeV state $^{16}$O]{
\includegraphics[scale=0.45]{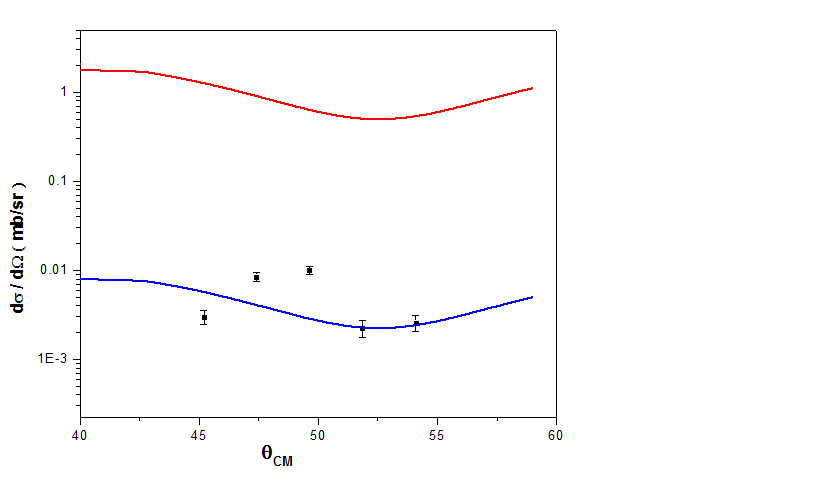}
}}
\mbox{
\subfigure[ 11.52 MeV state $^{16}$O]{
\includegraphics[scale=0.45]{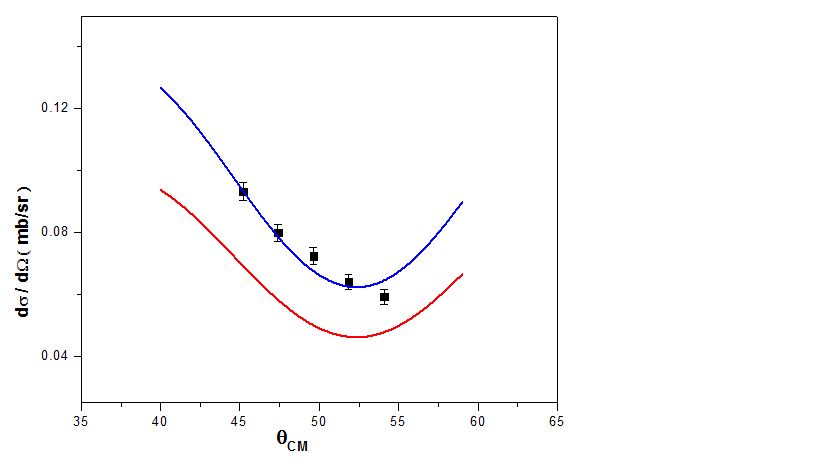}
}}
\caption{\label{fig.6}{\scriptsize Comparison of DWBA calculations using FRESCO of $^{12}$C($^{20}$Ne,$^{16}$O)$^{16}$O $\alpha$-transfer angular distribution of 6.92 MeV, 7.12 MeV and 11.52 MeV states of $^{16}$O with experimental data. The black square  are experimental data with error and solid red line is the FRESCO calculation and solid blue line is the FRESCO calculation after normalisation to the experimental data for (a) 6.92 MeV, (b) 7.12 MeV and (c) 11.52 MeV states of $^{16}$O respectively.}}
\label{subfig}
\end{figure*}


The $S_{\alpha}$ of the 6.92 MeV, 7.12 MeV and 11.52 MeV states are extracted by a normalization of the experimental angular distributions (in Fig.7) in terms of the calculated values using equation (1). The spectroscopic factor of $^{20}$Ne is adopted as 0.229 from [18]. The potential parameters used in the FRESCO calculations are shown in Table II. and Table III. The comparision of the calculations with data yields the $S_{\alpha}$ and the ANC values which are shown in Table IV. and Table V.

Error analysis of the extracted ANC and spectroscopic factors have been carried out. The sources of error in the spectroscopic factor are from experimental and theoretical DWBA cross-sections. The errors in the experimental cross-section are estimated from statistical and systematic errors. The statistical error is about 10$\%$. The source of systematic errors are from beam energy fluctuation, target thickness uncertainty, beam current fluctuation, solid angle uncertainty etc. and is estimated to be 5$\%$. The errors (statistical and systematic) are added in quadrature assuming them to be independenet errors to obtain the error in the spectroscopic factor. The  source of the theoretical errors in the DWBA calculation are from the uncertainty in the entrance, exit and core-core potentials.
The uncertainty in the ANC depends on the error in the spectroscopic factor and that in the single particle ANC (\textit{b}). The two errors are added in quadrature to obtain the error in the extracted ANC value. The error in single particle ANC (\textit{b}) is about 14$\%$.  
Fig.8  shows the $S_{E2}$ curve with the $^{12}$C($\alpha$,$\gamma$) E2 capture data [23,24] have been described by a four level R-matrix fit (6.92 MeV sub-threshold state, 9.85 MeV state, 11.52 MeV state and a higher background equivalent state) assuming a radius $R_{0}$= 6.5 fm for the inner space using the AZURE2 code [25]. The fitting is carried out with all parameters fixed except the energy and width of the background state which are considered to be freely varying. The ANC  of the 6.92 MeV  state was fixed at  (1.034 $\pm$ 0.4)$\times$ $10^{5}$ $fm^{-1/2}$ which is determined through the  $^{12}$C($^{20}$Ne,$^{16}$O)$^{16}$O reaction in the present case. Its energy and $\Gamma_{\gamma}$ have fixed to the values $E_{r}$= 0.2448 MeV and $\Gamma_{\gamma}$= 97 meV [26]. For the resonance parameters of the $E_{x}$= 9.85 MeV and $E_{x}$= 11.52 MeV states, we have used the value given in [27]. All the resonance parameters used in the R-matrix fit for the astrophysical S factor of the E2 component are given in the Table VI. The extracted $S_{E2}$ value from fig.8 at 300 keV is 52 $\pm$ 25 keV b. Comparision of astrophysical S factors at 300 keV obtained in various experiments including this work for E2 components are given in the Table VII.

 \begin{table*}[!htbp]
\caption{The potential parameters obtained from the elastic $^{20}$Ne + $^{12}$C scattering at $E_{lab}$=150 MeV in optical model analysis.\\} 
\centering 
\begin{tabular}{c c c c c c c c} 
\hline\hline  \\
System & V$_o$ (MeV) & r$_o$ (fm)  & a$_o$ (fm) & W (MeV) & r$_w$ (fm) & a$_w$ (fm) & Ref. \\ [0.6ex] 
\hline \\

$^{20}$Ne + $^{12}$C (WS) & 85.00 & 1.383  & 0.63 & 10.65 & 1.5 & 0.56 & [This work]  \\ [2ex] 
\hline 
\end{tabular}
\label{table:nonlin} 
\end{table*}

\begin{table*}[!htbp]
\caption{The potential parameters required in the calculation are
shown.} 
\centering 
\footnotesize
\begin{tabular}{c c c c c c c c c c} 
\hline\hline   \\
System & V$_o$ (MeV) & r$_o$ (fm)  & a$_o$ (fm) & W (MeV) & r$_w$ (fm) & a$_w$ (fm) & R$_{o}$ (fm) & r$_c$ (fm) & Ref. \\ [0.6ex] 
\hline \\ 
$^{20}$Ne + $^{12}$C (WS) & 85.00 & 1.383  & 0.63 & 10.65 & 1.5 & 0.56 & - & 1.25 & [This work]  \\  [2ex] 

$^{16}$O + $^{16}$O (WS) & 135.0 & 1.73  & 0.835 & 27.0 & 2.3 & 0.6 & - & 1.25 & [6]  \\ [2ex]

$^{16}$O + $^{12}$C (WS) & 282.2 & 0.586  & 0.978 & 13.86 & 1.183 & 0.656 & - & - & [16]  \\ [2ex]

$\alpha$ + $^{12}$C (WS) & 85.9  & 0.916  & 2.7 & - & - & - & - & - & [17]  \\ [2ex]

$\alpha$ + $^{16}$O (WS) & 13.496 & 2.303  & 5.4 & - & - & - & - & - & [-]  \\ [2ex] 
\hline 
\end{tabular}
\label{table:nonlin} 
\end{table*}

\begin{table*}[!htbp]
\caption{Comparison of spectroscopic factors of the $^{16}$O states deduced from our calculation with earlier works.} 
\centering 
\footnotesize
\begin{tabular}{c c c c c c c c} 
\hline\hline  \\
State of $^{16}$O (MeV) & Present work & S. Adhikari [5] & Oulebsir [19] & Bellhout [20] & Bechetti  & Keely [7] & Cobern [8]  \\ [0.6ex]  \\
\hline  \\
6.92 & 0.786 $\pm$ 0.56 & 0.32 $\pm$ 0.07  & 0.15 $\pm$ 0.05 & 0.37 $\pm$ 0.11  & 1.35 [9] and 4.134 [10] & 0.68 & 1.10  \\ [2ex] 
7.12 & 0.0196 $\pm$ 0.017 & 0.22 $\pm$ 0.07 & 0.07 $\pm$ 0.03 & 0.11 $\pm$ 0.03  &  $0.08^{+0.04}_{-0.06}$ [10] & - & 0.2  \\ [2ex]
11.52 & 5.89 $\pm$ 1.56 & - & - & -  & - & - & -  \\ [2ex] 
\hline 
\end{tabular}
\label{table:nonlin} 
\end{table*}

\begin{table*}[!htbp]
\caption{Comparison of ANC of 6.92 MeV  and 7.12 MeV states of $^{16}$O ontained from present work with earlier works.} 
\centering 
\footnotesize
\begin{tabular}{c   c    c} 
\hline\hline  \\

Experiment & $C^{2}$ ($2^{+}$) ($10^{10}$ $fm^{-1}$) & $\hspace{0.5cm}$ $C^{2}$ ($1^{-}$) ($10^{28}$ $fm^{-1}$)  \\ [0.2ex]  \\
\hline  \\
This work & ( 1.07 $\pm$ 0.16 ) & ( 6.56 $\pm$ 1.45 )  \\ [2ex]
A. Mondal [21,22] & 1.14 & 76.8  \\ [2ex]
S. Adhikari [5] & (3.0 $\pm$ 0.122) & ( 303.8 $\pm$ 22.09 )  \\ [2ex]
Brune [2] & (1.29 $\pm$ 0.23) & ( 4.33 $\pm$ 0.84 )  \\ [2ex]
Avila [12] & (1.48 $\pm$ 0.16) & ( 4.39 $\pm$ 0.59 )  \\ [2ex]
Belhout [20] & ($1.96^{+1.41}_{-1.27}$) & ( 3.48 $\pm$ 2.0 )  \\ [2ex]
Oulebsir [19] & (2.07 $\pm$ 0.8) & ( 4.0 $\pm$ 1.38 )  \\ [2ex]
\hline 
\end{tabular}
\label{table:nonlin} 
\end{table*}

\begin{figure*}[!htbp]
\centering
\includegraphics[scale=0.58]{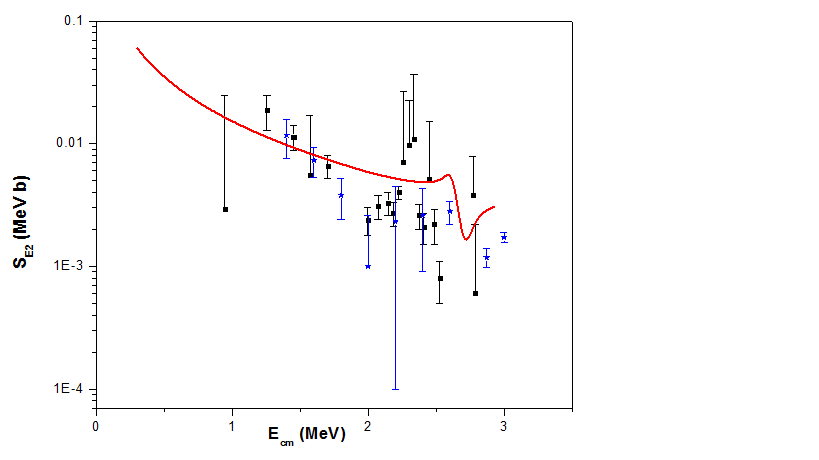}
\caption{\label{fig.1}{\scriptsize Astrophysical S factor for the $^{12}$C($\alpha$,$\gamma$)$^{16}$O reaction with R-matrix calculations of the E2 component. The solid square and solid star points are the experimental data from [23] and [24] respectively. The solid red line is the R-matrix fit using  ANC value of the 6.92 MeV state determined in the present work. }}
\end{figure*}

\begin{table*}[!htbp]
\caption{Resonance parameters used in the R-matrix fit of the astrophysical S factor of the E2 component. The values in the brackets are the fixed resonance parameters.} 
\centering 
\small
\begin{tabular}{c  c  c  c  c} 
\hline\hline  \\
$J^{\pi}$  & $E_{x}$ (MeV)   & $E_{r}$ (MeV)   & \hspace{0.4cm} ANC (C) in $fm^{-1/2}$ or $\Gamma_{\alpha}$ (keV)   & \hspace{0.4cm} $\Gamma_{\gamma}$ (keV) \\ [2ex]  \\
\hline  \\
$2^{+}$ & 6.92 & [-0.244]  & C=[(1.228 $\pm$ 0.6)$\times$ $10^{5}$] & [9.7$\times$ $10^{-5}$]  \\ [2ex]  
$2^{+}$ & 9.85 & [2.683] & $\Gamma_{\alpha}$=[0.76] & [5.7$\times$ $10^{-6}$]  \\ [2ex]
$2^{+}$ & 11.52 & [4.339] & $\Gamma_{\alpha}$=[83.0] & [6.1$\times$ $10^{-4}$]  \\ [2ex]
$2^{+}$ & Background & 7.84 & $\Gamma_{\alpha}$= 0.257 & 1.3$\times$ $10^{-4}$   \\ [2ex]
\hline 
\end{tabular}
\label{table:nonlin} 
\end{table*}

\begin{table*}[]
\caption{Comparision of astrophysical S factors at 300 keV obtained in various experiments including this work for E2 components.} 
\centering 
\small
\begin{tabular}{c  c  c  c  c  c  c  c  c} 
\hline\hline  \\
S factor  & This Work & Oulebsir[19]  & Brune[2]  & Tischauser[27] & Hammer[28] & Kunz[23] & NACRE[29] & Ouellet[24] \\
(0.3 MeV)& & & & & & & & \\
\hline  \\
$S_{E2}$ (keV b) & 52 $\pm$ 25 & 50 $\pm$ 19  & $44^{+19}_{-23}$  & 53 $\pm$ 13 & 81 $\pm$ 22 & 85 $\pm$ 30 & 120 $\pm$ 60 & 36 $\pm$ 6 \\ [2ex]  
\hline 
\end{tabular}
\label{table:nonlin} 
\end{table*}

\section{Conclusions}
In this experiment we have measured $^{12}$C($^{20}$Ne,$^{16}$O)$^{16}$O angular distribution for the population of $\alpha$-transfer states at $E_{lab}$=150 MeV. The ANC of the 6.92 MeV and 7.12 MeV states of $^{16}$O are determined for the first time using this reaction. The extraction of ANC were done by using the direct reaction code FRESCO. The required entrance channel potential of $^{20}$Ne+$^{12}$C was determined from the elastic scattering data measured in the same experiment. We have extracted $S_{E2}$ value at 300 keV by R-matrix calculation using ANC value of 6.92 MeV state which is determined from this reaction.

\section{Acknowledgement}
The authors acknowledge the help and support of all VECC Cyclotron staff, Jonaki Panja and Taniya Basu for the experiment.  \\ 

\end{document}